\documentclass[
amsmath,amssymb]{revtex4}
\usepackage{epsfig}
\usepackage{amsthm}
\usepackage{amssymb}
\usepackage{latexsym}
\usepackage{url}
\usepackage{amsmath, amscd}
\usepackage{verbatim}

\newtheorem{thm}{Theorem}

\theoremstyle{definition}

\newcommand{\R}{\mathbb{R}}

\begin{document}
\title{Functional methods underlying classical mechanics, relativity and quantum theory}
\author{A. \surname{Kryukov}} 
\affiliation{Department of Mathematics, University of Wisconsin Colleges, 780 Regent Street, Madison, Wisconsin 53708, USA} 
 
\begin{abstract}

The paper investigates the physical content of a recently proposed mathematical framework 
 that unifies the standard formalisms of classical mechanics, relativity and quantum theory. In the framework states of a classical particle are identified with Dirac delta functions. The classical space is "made" of these functions and becomes a submanifold in a Hilbert space of states of the particle. The resulting embedding of the classical space into the space of states is highly non-trivial and accounts for numerous deep relations between classical and quantum physics and relativity. One of the most striking results is the proof that the normal probability distribution of position of a macroscopic particle (equivalently, position of the corresponding delta state within the classical space submanifold) yields the Born rule for transitions between arbitrary quantum states.   
\end{abstract}


\maketitle

\section{Introduction}

Most of the modern theoretical physics is built on the following well accepted and experimentally verified theories: non-relativistic and relativistic classical mechanics, special relativity and Maxwell's electrodynamics, Einstein theory of gravity, quantum mechanics, quantum electrodynamics and gauge field theory. Despite the fact that each one of these theories embraces a wide range of phenomena, their interrelationship is far from being clear and leads to the well known difficulties and paradoxes. For instance, the special role played by the classical mechanical devices in quantum theory is a basis of the so-called measurement problem. The classical meaning of time in quantum mechanics makes it difficult to merge this theory with special relativity. The relationship of the quantum field theory and quantum mechanics is also far from being understood. General relativity so far has resisted all attempts of being quantized.  

To thoroughly understand the relationship between these theories it is necessary to find a common mathematical language underlying them. Currently the classical mechanics, classical gauge theory and the theory of gravity employ the language of differential geometry of finite dimensional differential manifolds with a metric and possible additional structures, and fibre bundles over such manifolds with a connection. On the other hand, mathematics of the quantum theory is based on algebras of linear operators in Hilbert spaces and the theory of representations of groups.  
It is expected that, when properly understood, quantum theory will contain the classical theory in a certain limit. 
If so, then the problem of finding a common language for the theories is, loosely speaking, the problem of encoding the non-linear theory of finite dimensional manifolds into (an extension of) the linear theory of infinite-dimensional Hilbert spaces and linear operators on them. 

A possible solution to the problem was recently proposed and investigated in Refs.\cite{KryukovIJMMS}-\cite{KryukovFOP}. The proposed formalism combines the strength of the functional analysis in Hilbert spaces, the theory of generalized functions, and differential geometry of Banach manifolds. It allows one to identify the classical space and space-time with submanifolds in a Hilbert space of quantum states. The classical mechanics and relativity theory are then reformulated in functional terms. The reformulation yields a new relativistic framework for quantum mechanics.  In the process quantum observables are identified with vector fields on the space of states. Quantum commutators become the Lie brackets of the fields. Other important physically meaningful identifications follow.

The goal of this paper is to demonstrate that the classical particle mechanics, relativistic mechanics, dynamics of particles in gravitational field and quantum mechanics can be all formulated in a very elegant way using the new formalism and the principle of least action.  The obtained unified treatment is then applied to shed new light into the nature of quantum uncertainty and the problem of the relationship between classical and quantum phenomena.


\section{Classical mechanics in Hilbert space}


The state of a spinless particle with a known position ${\bf a}$ in the Euclidean space $ \R^{3}$ is described in quantum mechanics by the delta function $\delta^{3}_{\bf a}({\bf x})=\delta^{3}({\bf x}-{\bf a})$. The map $\omega: {\bf a} \longrightarrow \delta^{3}_{\bf a}$ provides a one-to-one correspondence between points ${\bf a}$ and states $\delta^{3}_{\bf a}$.
In the single-particle case the classical space $\R^{3}$ will be then identified with the set of all delta functions in the space of state functions of the particle. To begin with, we need a Hilbert space that contains delta functions.
 
Consider the usual Hilbert space $L_{2}({\R}^{3})$ of Lebesgue square-integrable functions on $\R^{3}$. Note that the inner product of functions $\varphi, \psi \in L_{2}(\R^{3})$ can be formally written as follows:
\begin{equation}
\label{innerdd}
(\varphi, \psi)_{L_{2}}=\int \delta^{3}({\bf x}-{\bf y})\varphi({\bf x}){\overline \psi}({\bf y})d^{3}{\bf x}d^{3}{\bf y}.
\end{equation}
In particular, the fact that delta functions are not in $L_{2}(\R^{3})$ is related to the singularity of the kernel $\delta^{3}({\bf x}-{\bf y})$ of the Hilbert metric. We can approximate this kernel by the Gaussian function $\left(\frac{L}{\sqrt {2\pi}}\right)^{3}e^{-\frac{L^{2}}{2}({\bf x}-{\bf y})^{2}}$ for some positive constant $L$. This yields the product
\begin{equation}
(\varphi, \psi)_{\bf H}=\left(\frac{L}{\sqrt {2\pi}}\right)^{3}\int e^{-\frac{L^{2}}{2}({\bf x}-{\bf y})^{2}} \varphi({\bf x}){\overline \psi}({\bf y})d^{3}{\bf x}d^{3}{\bf y}.
\end{equation}
One can check that this is indeed an inner product on $L_{2}({\R}^{3})$. 
The Hilbert space ${\bf H}$ obtained by completing the space $L_{2}({\R}^{3})$ in this inner product contains delta functions  $\delta^{3}({\bf x}-{\bf a})$ and their derivatives. 
 Moreover, by choosing $L$ sufficiently large (or by choosing appropriate units), one can make the norm of any square-integrable function in this metric as close as desired to its  $L_{2}({\R}^{3})$-norm. This relationship will be denoted by ${\bf H} \approx L_{2}(\R^{3})$.
Choosing for now $L=1$ and dropping the coefficient $(1/{\sqrt {2\pi}})^{3}$ we obtain the product 
\begin{equation}
\label{hh}
(\varphi, \psi)_{\bf H}=\int e^{-\frac{1}{2}({\bf x}-{\bf y})^{2}}\varphi({\bf x}){\overline \psi}({\bf y})d^{3}{\bf x}d^{3}{\bf y}.
\end{equation}
Formally,
\begin{equation}
\int e^{-\frac{1}{2}({\bf x}-{\bf y})^{2}}\delta^{3}({\bf x}-{\bf a})\delta^{3}({\bf y}-{\bf a})d^{3}{\bf x}d^{3}{\bf y}=1,
\end{equation}
so that the norm of the delta function  $\delta^{3}({\bf x}-{\bf a})$ in the resulting space ${\bf H}$ is $1$.

Consider the set $M_{3}$ of all delta functions  $\delta^{3}_{\bf a}({\bf x})\equiv \delta^{3}({\bf x}-{\bf a})$ in ${\bf H}$. To know position ${\bf a}$ of a classical particle in $\R^{3}$ is to know the corresponding point $\delta^{3}_{\bf a}$ in $M_{3}$. Consider a path ${\bf r}={\bf a}(t)$ with values in $\R^{3}$ and the corresponding path $\varphi=\delta^{3}_{{\bf a}(t)}$ in $M_{3}$.
With the use of the chain rule the velocity vector $d \varphi/dt$ can be written as
\begin{equation}
\label{chain}
\frac{d \varphi}{dt}=-\frac{\partial}{\partial {\bf x}^{i}}\delta^{3}({\bf x}-{\bf a})\frac {d{\bf a}^{i}}{dt},
\end{equation}
where the usual summation convention for repeating indices is accepted.
It follows that the norm of the velocity in the space ${\bf H}$ is
\begin{equation}
\left \| \frac{d \varphi}{dt} \right \|^{2}_{H}=
\int k({\bf x},{\bf y}) \frac{\partial}{\partial x^{i}}\delta^{3}({\bf x}-{\bf a})\frac {d{\bf a}^{i}}{dt}\frac{\partial}{\partial y^{k}}\delta^{3}({\bf y}-{\bf a})
\frac {d{\bf a}^{k}}{dt}d^{3}{\bf x}d^{3}{\bf y},
\end{equation}
where $k({\bf x},{\bf y})=e^{-\frac{1}{2}({\bf x}-{\bf y})^{2}}$.
``Integration by parts" in the last expression gives
\begin{equation}
\label{parts1}
\left \| \frac{d \varphi}{dt} \right \|^{2}_{H}=
\left.\frac {\partial^{2}k({\bf x},{\bf y})}{\partial x^{i} \partial y^{k}}\right|_{{\bf x}={\bf y}={\bf a}} \frac {d{\bf a}^{i}}{dt}\frac {d{\bf a}^{k}}{dt}.
\end{equation}
Because 
\begin{equation}
\left.\frac {\partial^{2}k({\bf x},{\bf y})}{\partial x^{i} \partial y^{k}}\right|_{{\bf x}={\bf y}={\bf a}}=\delta_{ik},
\end{equation}
where $\delta_{ik}$ is the Kronecker delta symbol, we obtain equality of the speeds
\begin{equation}
\label{Norms}
\left \| \frac{d \varphi}{dt} \right \|_{H}=\left \| \frac{d {\bf a}}{dt} \right \|_{\R^{3}}.
\end{equation}
Accordingly, the set $M_{3}$ as a metric subspace of ${\bf H}$ is identical to the Euclidean space $\R^{3}$. Notice however that $M_{3}$ is not a vector subspace of ${\bf H}$. Rather, because the norm of delta functions in ${\bf H}$ is $1$,  the metric space $M_{3}$ is a submanifold of the unit sphere $S^{\bf H}$ in ${\bf H}$. Since delta functions $\delta^{3}_{{\bf a}_{k}}$ with different ${{\bf a}_{k}}$, $k=1,...,n$ are linearly independent, the manifold $M_{3}$ ``spirals'' through dimensions of the sphere, forming a complete subset of ${\bf H}$. This means that no function in ${\bf H}$ is orthogonal to the submanifold $M_{3}$. 

Nevertheless, a vector structure on $M_{3}$ exists. Namely, the operations of addition $\oplus$ and multiplication by a scalar $\lambda \odot$ can be defined via $\omega({\bf a})\oplus\omega({\bf b})=\omega({\bf a}+{\bf b})$ and $\lambda \odot\omega({\bf a})=\omega(\lambda {\bf a})$, where the map $\omega:\R^{3}\longrightarrow {\bf H}$ is given by $\omega:{\bf a} \longrightarrow \delta^{3}_{\bf a}$. The resulting operations are continuous in the topology of $M_{3}\subset {\bf H}$. That is, the metric space $M_{3}$ with this vector structure is isomorphic to the vector space $\R^{3}$ with the Euclidean metric. 

Now that a functional realization of the classical space $\R^{3}$ and of position ${\bf a}$  and velocity $\frac{d{\bf a}}{dt}$ of a material point in the space are found, the dynamics of the point can be derived from the principle of least action. Namely, consider the action functional
\begin{equation}
S=\int e^{-\frac{1}{2}({\bf x}-{\bf y})^{2}}\left[\frac{m}{2}\frac{d \varphi_{t}({\bf x})}{dt} \frac{d{\overline  \varphi_{t}}({\bf y})}{dt}-V(x) \varphi_{t}({\bf x}) {\overline \varphi_{t}}({\bf y})\right]d^{3}{\bf x}d^{3}{\bf y}dt,
\end{equation}
where $m$ is the mass of the particle, $V$ is the potential, and $\varphi_{t}$ is constrained to be on the submanifold $M_{3}\subset {\bf H}$, i.e., $\varphi_{t}({\bf x})=\delta^{3}({\bf x}-{\bf a}(t))$. Using (\ref{chain}) and integrating by parts as in (\ref{parts1}), we immediately obtain
\begin{equation}
S=\int\left[\frac{m}{2}\left(\frac{d{\bf a}}{dt}\right)^{2}-V({\bf a})\right]dt,
\end{equation}
i.e., the usual action functional for a material point in classical mechanics.

This demonstrates that the classical mechanics of a material point can be formulated in purely functional terms. Position and velocity of the point are now given by functions in a Hilbert space. The dynamics of the point is derived from the principle of least action by variation of paths in the Hilbert space. The condition that the position of the point at any moment of time is well defined is the constraint $\varphi_{t}({\bf x})=\delta^{3}({\bf x}-{\bf a}(t))$. 
The relation between $d\varphi_{t}/dt$ and $d{\bf a}/dt$ is given by (\ref{chain}) and (\ref{Norms}). 
That is, the velocity $d\varphi_{t}/dt$ of state in the particular case when the state travels along the submanifold $M_{3}$ gets identified with the usual velocity $d{\bf a}/dt$. 

Note that the momentum of the delta state has infinite dispersion when the $L_{2}(\R^{3})$ metric is used. The dispersion is finite in the ${\bf H}$ metric. Alternatively, the motion $\varphi_{t}({\bf x})=\delta^{3}({\bf x}-{\bf v}t)$ in ${\bf H}$ can be identified with the motion of a Gaussian packet with group velocity ${\bf v}$ in the space $L_{2}(\R^{3})$. The packet is not an eigenstate of the momentum operator, but its group velocity is well defined. The velocity $d\varphi_{t}/dt$ gets identified with the group velocity ${\bf v}$.

Note also that because state functions considered here are not the eigenstates of the momentum operator, a simultaneous description of position and velocity of the particles in the constructed realization does not contradict the uncertainty principle.  
Finally, as briefly discussed in section XII, systems of $n$ interacting particles can be considered in a similar way.

\section{Relativistic mechanics in Hilbert space}

A functional realization of the relativistic mechanics requires a space of functions of four variables ${\bf x}, t$. Let's begin with the Hilbert space $L_{2}(\R^{4})$ of square-integrable functions of four variables $x=({\bf x}, t)$ and complete it in the metric given by the kernel $e^{-\frac{1}{2}(x-y)^{2}}$. As before, the resulting Hilbert space ${\widetilde H}$ contains delta functions $\delta^{4}_{a}$ and the set $M_{4}$ of all delta functions forms a submanifold of ${\widetilde H}$ with the induced metric of the Euclidean space $\R^{4}$. 

To obtain the Minkowski space metric consider the Hermitian form $(f,g)_{H_{\eta}}$ given in the units where the speed of light $c$ is $1$ by 
\begin{equation}
\label{innerM}
(f,g)_{H_{\eta}}=\int e^{-\frac{1}{2}({\bf x}-{\bf y})^{2}+\frac{1}{2}(t-s)^{2}}f({\bf x},t)\overline{ g({\bf y},s)} d^{3}{\bf x}dtd^{3}{\bf y}ds.
\end{equation}
Let $(f,f)_{H_{\eta}}\equiv \left\|f\right\|^{2}_{H_{\eta}}$ be the corresponding quadratic form, or the squared $H_{\eta}$-norm. 
Because of the positive term in the exponent of the kernel in (\ref{innerM}), not every function in the Hilbert space ${\widetilde H}$ has a finite $H_{\eta}$-norm. In addition, the quadratic form $(f,f)_{H_{\eta}}$ is not positive-definite. More precisely, if $f\neq 0$ and $f({\bf x},t)$ is even in $t$, then $(f,f)_{H_{\eta}}>0$. Likewise, if $f\neq 0$ and $f({\bf x},t)$ is odd in $t$, then $(f,f)_{H_{\eta}}<0$. So, let $H$ be the set of functions in ${\widetilde H}$ whose even and odd in $t$ components have a finite $H_{\eta}$-norm. As shown in Ref.\cite{KryukovJMP1}, $H$ is exactly the set of all functions $f({\bf x},t)=e^{-t^{2}}\varphi({\bf x},t)$ with $\varphi \in \widetilde{H}$. Moreover, $H$ furnished with the inner product $(f,g)_{H_{+}}=(\varphi, \psi)_{\widetilde{H}}$, where $f({\bf x},t)=e^{-t^{2}}\varphi({\bf x},t)$, $g({\bf x},t)=e^{-t^{2}}\psi({\bf x},t)$, is a Hilbert space. The Hermitian form (\ref{innerM}) defines an indefinite, non-degenerate inner product on $H$. Finally, $H$ contains the delta functions $\delta^{4}_{a}(x)=\delta^{4}(x-a)$ and their derivatives for all $a \in \R^{4}$. 

The obtained Hilbert space $H$ with the additional indefinite metric is an example of what is called a {\em Krein space}. 
Let $M_{4}$ be the set of all delta functions in $H$ and let $N$ be the Minkowski space-time. 
Then, similarly to the case of the Euclidean space $\R^{3}$, the set $M_{4}$ is a submanifold of the Hilbert space $H$ that is diffeomorphic to $N$. Moreover, the indefinite metric on $H$ yields the Minkowski metric on $M_{4}$. 
In fact, consider a path $x=a(\tau)$ with values in $N$ and the corresponding path $\varphi=\delta^{4}_{a(\tau)}$ in $M_{4}$.
As in (\ref{chain}), the velocity vector $d \varphi/d\tau$ can be written as
\begin{equation}
\label{chain1}
\frac{d \varphi}{d\tau}=-\frac{\partial}{\partial x^{\mu}}\delta^{4}(x-a)\frac{da^{\mu}}{d\tau},
\end{equation}
where summation goes over $\mu=0,1,2,3$.
It follows that the norm of the velocity in the indefinite metric on the space $H$ is
\begin{equation}
\left \| \frac{d \varphi}{d\tau} \right \|^{2}_{H_{\eta}}=
\int k(x,y) \frac{\partial}{\partial x^{\mu}}\delta^{4}(x-a)\frac {da^{\mu}}{d\tau}\frac{\partial}{\partial y^{\nu}}\delta^{4}(y-a)
\frac {da^{\nu}}{d\tau}d^{4}xd^{4}y,
\end{equation}
where $k(x,y)=e^{-\frac{1}{2}(x-y)_{\eta}^{2}}$ and $\eta$ stands for the Minkowski norm.
``Integration by parts" in the last expression gives
\begin{equation}
\label{parts}
\left \| \frac{d \varphi}{d\tau} \right \|^{2}_{H_{\eta}}=
\left.\frac {\partial^{2}k(x,y)}{\partial x^{\mu} \partial y^{\nu}}\right|_{x=y=a} \frac {d a^{\mu}}{d\tau}\frac {da^{\nu}}{d\tau}
\end{equation}
and since 
$\left.\frac {\partial^{2}k(x,y)}{\partial x^{\mu} \partial y^{\nu}}\right|_{x=y=a} $ is equal to the Minkowski metric $\eta_{\mu\nu}$, we have
\begin{equation}
\left \| \frac{d \varphi}{d\tau} \right \|_{H_{\eta}}=\left \| \frac{d a}{d\tau} \right \|_{\eta},
\end{equation}
i.e., the equality of the speed of evolution in the Hilbert space $H$ to the usual $4$-speed. In particular, if $\tau$ is the proper time parameter,  $\left \| \frac{d \varphi}{d\tau} \right \|_{H_{\eta}}$ is equal to the speed of light (i.e., $1$ in the chosen units).

A motion of a macroscopic particle in relativistic mechanics is now realized as a motion in $M_{4}\subset H$. The dynamics of a free particle follow from the principle of the least action for the square-length action functional
\begin{equation}
\label{squarelength}
S=\frac{m}{2}\int k(x,y) \frac{d\varphi_{\tau}(x)}{d\tau} \frac{d{\overline \varphi}_{\tau}(y)}{d\tau}d^{4}xd^{4}yd\tau,
\end{equation}
 $k(x,y)=e^{-\frac{1}{2}(x-y)_{\eta}^{2}}$,
under the constraint $\varphi_{\tau}(x)=\delta^{4}(x-a(\tau))$. In fact, for $S$ with this constraint we have
\begin{equation}
S=\frac{m}{2}\int \left \| \frac{d a}{d\tau} \right \|^{2}_{\eta}d\tau,
\end{equation}
which is the correct action functional for a free relativistic particle of mass $m$.

\section{Gravity in Hilbert space}

As demonstrated in Ref.\cite{KryukovJMP2}, an arbitrary curved metric on space-time can be obtained in the described way. That is, for any Riemannian or pseudo-Riemannian metric $g$ on space time $N$ there exists a Krein space $H$ that contains delta functions $\delta^{4}_{a}$ and such that the metric induced on the set $M_{4}$ of these delta functions coincides with $g$, at least locally. 

It follows in particular that by an appropriate ``curving" of the metric (\ref{innerM}) in the action  (\ref{squarelength}) it should be possible to derive the law of motion of particles in the  field of gravity. In other words, geodesic motion in an arbitrary curved classical space-time can be identified with a geodesic motion on the submanifold $M_4$ with the induced metric.

To see how this can be done, consider the indefinite Hermitian metric with the kernel
\begin{equation}
\label{metricU}
k(x,y)=e^{-\frac{1}{2}({\bf x}-{\bf y})^{2}+\frac{1}{2}(1+u({\bf x})+u({\bf y}))(t-s)^{2}},
\end{equation}
where $u$ is an appropriate function (the symmetric sum $u({\bf x})+u({\bf y})$ is needed for Hermicity of the resulting $2$-form). Note that when $u=0$ we get the already familiar metric. 
Under the constraint $\varphi_{\tau}(x)=\delta^{4}(x-a(\tau))$ the square-length action functional (\ref{squarelength}) with this metric yields
\begin{equation}
\label{Newton}
S=\frac{m}{2}\int k(x,y) \frac{d\varphi_{\tau}(x)}{d\tau} \frac{d{\overline \varphi}_{\tau}(y)}{d\tau}d^{4}xd^{4}yd\tau=\frac{m}{2}\int \left.\frac {\partial^{2}k(x,y)}{\partial x^{\mu} \partial y^{\nu}}\right|_{x=y=a} \frac {d a^{\mu}}{d\tau}\frac {da^{\nu}}{d\tau}d\tau=\frac{m}{2}\int g_{\mu\nu} \frac {d a^{\mu}}{d\tau}\frac {da^{\nu}}{d\tau}d\tau,
\end{equation}
where $g_{\mu\nu}$ is given by the length element  $ds^{2}=(1+2u({\bf x}))dt^{2}-d{\bf x}^{2}$. As is well known, Einstein equations for the metric $ds^{2}$ in the non-relativistic approximation yield the Newton's law of motion of the particle in gravitational potential $u$. In other words, variation of the functional (\ref{Newton}) gives the equation
\begin{equation}
\frac{d{\bf v}}{dt}=-\nabla u.
\end{equation}

\section{Quantum mechanics in Hilbert space with indefinite metric}

Recall that in Sec.III the motion of a material point in relativistic mechanics was derived 
by variation of the square-length functional in a Hilbert space $H$ with indefinite metric (Krein space), defined in Sec.III. An additional constraint was the condition that the particle has a well defined position at any time. In other words, the path of the particle in the space of states $H$ has the form $\varphi_{\tau}( x)=\delta^{4}(x-a(\tau))$. Let's show that the non-relativistic quantum mechanics can be derived in a similar way, by dropping the condition of definiteness of the position, but preserving the condition of definiteness of time.  
Namely, we begin with the relativistic framework offered by the Krein space $H$, impose the constraint  $\varphi_{\tau}({\bf x},t)=\psi({\bf x},t)\delta(t-\tau)$ on the path of a particle and derive the usual non-relativistic quantum mechanics of the particle.

To do this let's first relate the Hilbert space $H$ of functions of four variables ${\bf x}, t$ with metric (\ref{innerM}) to the usual Hilbert spaces of functions of three variables ${\bf x}$ with $t$ as a parameter of evolution. For this consider the family of subspaces $H_{\tau}$ of $H$ each consisting of all functionals $\varphi_{\tau}({\bf x},t)=\psi({\bf x},t)\delta(t-\tau)$ for some fixed $\tau \in \R$. The inner product of any two functionals in $H_{\tau}$ in either ${\widetilde H}$ or $H_{\eta}$ metrics of Sec. III is simply the inner product in the space ${\bf H}$ introduced in Sec.II. This is the case because the factor $\delta(t-\tau)$ eliminates integration in $t$ and makes the term $(t-s)^{2}$ in the exponent of the kernel of the metric vanish. In the following, $H_{\tau}$ will be understood as a Hilbert space with this inner product. 
The map  $I:H_{\tau}\longrightarrow {\bf H}$ defined by $I(\varphi_{\tau})({\bf x})=\psi({\bf x}, \tau)$  is then an isomorphism of Hilbert spaces. 

Because ${\bf H} \approx L_{2}(\R^{3})$, the map $I$ basically identifies each subspace $H_{\tau}$ with the usual space $L_{2}(\R^{3})$ of state functions on $\R^{3}$ considered at time $\tau$. The reason why these particular subspaces are physically meaningful becomes clear from the following result that relates the dynamics on the family of subspaces $H_{\tau}$ and the usual space $L_{2}(\R^{3})$ of states of a spinless non-relativistic particle. 
\begin{thm}
\label{t1}
Let ${\widehat h}=-\Delta+V({\bf x},t)$ be the usual Hamiltonian of non-relativistic quantum mechanics of a single particle. Then the function $\psi({\bf x},t)$ satisfies the Schr{\"o}dinger equation $\frac{\partial \psi({\bf x},t)}{\partial t}=-i{\widehat h}\psi({\bf x},t)$ if and only if the path $\varphi_{\tau}({\bf x},t)=\psi({\bf x},t)\delta(t-\tau)$ in $H$ satisfies the equation $\frac{d\varphi_{\tau}}{d\tau}=\left(-\frac{\partial}{\partial t}-i{\widehat h}\right)\varphi_{\tau}$. 
\end{thm}
More generally, the result holds true when $\varphi_{\tau}({\bf x},t)=\psi({\bf x},t){\tilde \delta(t-\tau)}$ for some function ${\tilde \delta}$ of $t-\tau$. In the following ${\tilde \delta}$ will be taken to be a Gaussian function that approximates delta function.

Note that an equation of the form $\frac{d\varphi_{\tau}}{d\tau}=\left(-\frac{\partial}{\partial t}-i{\widehat h}\right)\varphi_{\tau}$ is used in Floquet theory that deals with Hamiltonians periodic in time Ref.\cite{Floquet} and in problems with time-dependent Hamiltonians Ref.\cite{Holand}.
It is also a well known non-relativistic limit of the Stueckelberg-Schr{\"o}dinger equation in the theory of Stueckelberg Ref.\cite{Stu1} and Horwitz \& Piron Ref.\cite{HorPir}. This theory treats space and time symmetrically and predicts interference in time Refs.\cite{Hor},\cite{Hor2}.
The non-relativistic limit of Stueckelberg theory was investigated by Horwitz and Rotbart Ref.\cite{HorRot}. The approximate equality of the time variable $t$ with the evolution parameter $\tau$ obtained in Ref.\cite{HorRot} is consistent with the definition of $H_{\tau}$.

\section{Arc-length action functional for quantum dynamics} 

Dynamics of a classical particle in a gravitational field follow from the variational principle for the arc length action functional. The resulting path is a geodesic.  
The following result demonstrates that the Schr{\"o}dinger dynamics can be derived in a similar fashion.

\begin{thm}
\label{hsquare}
Suppose that the evolution of a system satisfies the equation $\frac{d\varphi_{\tau}}{d\tau}=-i{\widehat A}\varphi_{\tau}$, where ${\widehat A}$ is an invertible self-adjoint operator in the space $L_{2}$ of Lebesgue square-integrable functions on a set. Then the operator ${\widehat K}={\widehat A}^{-1}\left({\widehat A}^{-1}\right)^{\ast}={\widehat A}^{-2}$ defines an inner product on the image $\mathrm{R}({\widehat A})$ of ${\widehat A}$. Let $H$ be the Hilbert completion of $\mathrm{R}({\widehat A})$. Then ${\widehat A}$ considered as a map into $H$ is bounded and  solutions of the evolution equation are geodesics on the sphere $S^{g}$ of unit-normalized states in $L_{2}$ with Riemannian metric $g$ defined by the inner product in $H$.
\end{thm}

For a finite dimensional example consider the space of spin states of a non-relativistic electron. Suppose the Hamiltonian is given by ${\widehat h}_{M}=M-\mu{\widehat \sigma}\cdot {\bf B}$, where ${\bf B}$ is  a homogeneous magnetic field, $\mu$ is the electron's magnetic moment, ${\widehat \sigma}$ is given by Pauli matrices ${\widehat \sigma}=(\sigma_{1}, \sigma_{2}, \sigma_{3})$ and $M=M\cdot I$ is a constant, such that $M \gg \mu B$.

The metric on the sphere of states $S^{3}$ in which solutions to the Schr{\"o}dinger equation
\begin{equation}
\label{newSchroed}
\frac{d \psi_{t}(s)}{dt}=-i{\widehat h}_{M}\psi_{t}(s).
\end{equation}
 are geodesics is given the operator ${\widehat h}_{M}^{-2}$. Because $\left(\mu{\widehat \sigma}\cdot {\bf B}\right)^{2}=\mu^{2}B^{2}$, where $B$ is the norm of ${\bf B}$, we have:
\begin{equation}
{\widehat h}_{M}^{2}=(M-\mu{\widehat \sigma}\cdot {\bf B})^{2}=M^{2}+\mu^{2}B^{2}-2M\mu{\widehat \sigma}\cdot {\bf B}.
\end{equation}
Suppose that the field ${\bf B}$ is directed along the $Z$-axis. Then ${\widehat \sigma}\cdot {\bf B}={\widehat \sigma_{z}}B$ and
\begin{equation}
{\widehat h}_{M}^{2}=
\left[ 
\begin{array}{cc}
\left(M-\mu B\right)^{2} &  0 \\ 
0 & \left(M+\mu B\right)^{2}
\end{array}
\right].
\end{equation}
Accordingly, the metric operator ${\widehat h}_{M}^{-2}$ has the form:
\begin{equation}
\label{hm}
{\widehat h}_{M}^{-2}=
\left[ 
\begin{array}{cc}
\frac{1}{\left(M-\mu B\right)^{2}} &  0 \\ 
0 & \frac{1}{\left(M+\mu B\right)^{2}}
\end{array}
\right].
\end{equation}

Consider now the metric $G$ induced on the unit sphere $S^{3}$ embedded into the space $C^{2}$ with metric given by (\ref{hm}). 
Since $M \gg \mu B$, this metric almost consides with the standard metric on the sphere of radius $1/M$. More precisely, the sphere with the metric $G$ is a $3$-ellipsoid
\begin{equation}
\frac{x_{1}^{2}}{1/(M-\mu B)^{2}}+\frac{y_{1}^{2}}{1/(M-\mu B)^{2}}+\frac{x_{2}^{2}}{1/(M+\mu B)^{2}}+\frac{y_{2}^{2}}{1/(M+\mu B)^{2}}=1
\end{equation}
in the Euclidean space $\R^{4}=C^{2}$.

Solutions to the Schr{\"o}dinger equation (\ref{newSchroed}) are geodesics on the ellipsoid and are given by
\begin{equation}
\label{newev}
\psi_{t}=
\left[ 
\begin{array}{cc}
\psi_{0+}e^{-i(M+\mu B)t} ,& \psi_{0-}e^{-i(M-\mu B)t} 
\end{array}
\right],
\end{equation}
where $[\psi_{0+},\psi_{0-}]$ is the initial spin state.
The Schr{\"o}dinger evolution is a slightly deformed phase motion. The curve (\ref{newev}) ``spirals'' along the sphere of states. In particular, after the time interval $[0, \frac{2\pi}{M}]$ the curve that starts at the point $[\psi_{0+}, \psi_{0-}]$ goes around the sphere and comes to a neighboring point 
\begin{equation}
\left[ 
\begin{array}{cc}
\psi_{0+}e^{-i\frac{2\pi \mu B}{M}} ,& \psi_{0-}e^{i\frac{2\pi \mu B}{M}} 
\end{array}
\right].
\end{equation}
Notice that the deformation of the metric on the sphere of states in the example is due to the interaction of the electron with the field.

It is also possible to find the length-type action functional on functions of four variables that yields the Schr{\"o}dinger evolution. The idea is to apply Theorem \ref{hsquare} to the case of the self-adjoint operator ${\widehat A}=-i\frac{\partial}{\partial t}+{\widehat h}$, acting in the space $L_{2}(\R^{4})$ of square-integrable functions of ${\bf x},t$ (see Theorem \ref{t1}). 
At first sight it seems that this will not work, because the kernel of ${\widehat A}$ is non-trivial. More precisely, the kernel consists of all solutions to the Schr{\"o}dinger equation. However, it is easy to see that those solutions are not in the space $L_{2}(\R^{4})$. In fact, since the  Schr{\"o}dinger evolution is unitary, the $L_{2}(\R^{3})$-norm of solutions is preserved in time, so that the $L_{2}(\R^{4})$-norm of any non-trivial solution is infinite. It follows that the operator ${\widehat A}$ is invertible and the Theorem \ref{hsquare} applies.  The result is the  following theorem
\begin{thm}
Let $H$ be the Hilbert space in the Theorem \ref{hsquare} and suppose that  the functions  $\varphi_{\tau}({\bf x},t)=\psi({\bf x},t){\tilde \delta}(t-\tau)$, $\psi({\bf x},t)\in L_{2}(\R^{3})$ are in $H$. 
Then solutions to the Schr{\"o}dinger equation are extrema of the square-length action functional
$S=\int_{a}^{b} \left({\widehat K} \frac{d\varphi_{\tau}}{d\tau}, \frac{d\varphi_{\tau}}{d\tau}\right)d\tau$,
subject to the constraint $\varphi_{\tau} \in S^{g}$, $\varphi_{\tau}({\bf x},t)=\psi({\bf x},t){\tilde \delta}(t-\tau)$, where $S^{g}$ is the unit sphere in $L_{2}(\R^{4})$ with the Riemannian metric given by the metric ${\widehat K}={\widehat A}^{-2}$ on $H$.
\end{thm}

Note that the operator ${\widehat A}^{-1}$ considered on functions $\varphi_{\tau}({\bf x},t)=\psi({\bf x},t)\delta(t-\tau)$ acts as the Green function of the Schr{\"o}dinger equation. In fact, if $x=({\bf x},t)$, $y=({\bf y},s)$ and $g(x;y)$ is the kernel of ${\widehat A}^{-1}$, then
\begin{equation}
\int g(x;y)\varphi_{\tau}(y)d^{4}y=\int g({\bf x},t; {\bf y},\tau)\psi({\bf y}, \tau)d^{3}{\bf y}
\end{equation}
and
\begin{equation}
\left(-\frac{\partial}{\partial t}-i{\widehat h}\right)g({\bf x},t; {\bf y},\tau)=\delta^{3}({\bf x}-{\bf y})\delta(t-\tau).
\end{equation}
So the Riemannian metric that makes Schr{\"o}dinger evolution on the unit sphere in  $L_{2}(\R^{4})$ a geodesic motion is given by a propagator of the Schr{\"o}dinger equation. Note that the free particle propagator
\begin{equation}
\label{p2}
g({\bf x}, t; {\bf y}, s)=\left(\frac{m}{2\pi i(t-s)}\right)^{\frac{3}{2}}e^{-\frac{m({\bf x}-{\bf y})^{2}}{2i(t-s)}},
\end{equation}
has the exponential form similar to the metric (\ref{hh}) in the space ${\bf H}$. In particular, the resulting space contains delta functions and the metric induced on the three-dimensional manifold $M_{3}$ of all delta functions is Euclidean.

\section{Vector representation in quantum mechanics}

Quantum evolution is a path on the unit sphere of states in a Hilbert space $L_{2}$ (modulo the domain issues of involved operators). Given a self-adjoint operator ${\widehat A}$ on  $L_{2}$, consider the vector field $A_{\varphi}=-i{\widehat A} \varphi$. This field is tangent to the sphere, because its integral curves are given by the family of unitary transformations applied to a point on the sphere: $\varphi_{t}=e^{-i{\widehat A}} \varphi_{0}$, so that the curves lie on the sphere.  There  is a simple relationship between the commutator of operators and the Lie bracket of the associated vector fields:
\begin{equation}
\label{comm}
[A_{\varphi},B_{\varphi}]=[{\widehat A},{\widehat B}]\varphi.
\end{equation}
Furthermore, a Hilbert metric on the space of states yields a Riemannian metric on the sphere. For this consider the realization $L_{2\R}$ of the Hilbert space $L_{2}$, i.e., the real vector space of pairs $X=({\mathrm Re} \psi, {\mathrm Im} \psi)$ with $\psi$ in $L_{2}$. If $\xi, \eta$ are vector fields on $S^{L_{2}}$, define a Riemannian metric $G_{\varphi}: T_{\R\varphi}S^{L_{2}}\times T_{\R\varphi}S^{L_{2}} \longrightarrow \R$ on the sphere by
\begin{equation}
\label{Riem}
G_{\varphi}(X,Y)={\mathrm Re} (\xi, \eta).
\end{equation}
Here $X=({\mathrm Re} \xi, {\mathrm Im} \xi)$, $Y=({\mathrm Re} \eta, {\mathrm Im} \eta)$ and  $(\xi, \eta)$ denotes the $L_{2}$-inner product of $\xi, \eta$. 

The value of the Lie bracket of vector fields associated with observables, that is, the commutator of observables, is related to the sectional curvature of the obtained Riemannian metric on $S^{L_{2}}$, Ref.\cite{KryukovFOP1}. The usual commutators are obtained when all sectional curvatures are equal to one in the corresponding Planck units. In such a way quantum non-commutativity gets encoded into the geometry of the sphere of states.
Note that the minuscule size of the sphere $S^{\bf H}$ in spatial (i.e., tangent to $M_{3}$) directions does not contradict the fact that $M_{3}$ itself is the infinite Euclidean space. This is because $M_{3}$ gets large by "spiraling" through dimensions of ${\bf H}$, which has nothing to do with the size of the sphere.

The Riemannian metric on $S^{L_{2}}$ yields a Riemannian (Fubini-Study) metric on the projective space $CP^{L_{2}}$, which is the base of the fibration $\pi:S^{L_{2}} \longrightarrow CP^{L_{2}}$. For this an arbitrary tangent vector $X \in T_{R\varphi}S^{L_{2}}$ is decomposed into two components: tangent and orthogonal to the fibre $\{\varphi\}$ through $\varphi$ (i.e., to the plane $C^{1}$ containing the circle $S^{1}=\{\varphi\}$). The differential $d\pi$ maps the tangential component to zero-vector. The orthogonal component of $X$ can be then identified with $d\pi(X)$. 
If two vectors $X,Y$ are orthogonal to the fibre $\{\varphi\}$, the inner product of $d\pi(X)$ and $d\pi(Y)$ in the Fubini-Study metric is equal to the inner product of $X$ and $Y$ in the metric $G_{\varphi}$:
\begin{equation}
(d\pi(X), d\pi(Y))_{FS}= G_{\varphi}(X,Y).
\end{equation}

Using this, it is now easy to interpret geometrically the expected value and the uncertainty of an observable, Ref.\cite{KryukovUncert}. Namely,
 the equality
\begin{equation}
{\overline A} \equiv (\varphi, {\widehat A}\varphi)=(-i\varphi, -i{\widehat A}\varphi),
\end{equation}
(with the usual $L_{2}$ inner product) signifies that the expected value of an observable ${\widehat A}$ in the state $\varphi$ is the projection of the vector $-i{\widehat A}\varphi \in T_{\varphi}S^{L_{2}}$ on the vector $-i\varphi=-i I \varphi \in T_{\varphi}S^{L_{2}}$, associated with the identity operator $I$ and tangent to the phase circle through $\varphi$. 
Because
\begin{equation}
(\varphi, {\widehat A}^{2}\varphi)=({\widehat A}\varphi, {\widehat A}\varphi)=(-i{\widehat A}\varphi, -i{\widehat A}\varphi),
\end{equation}
the term $(\varphi, {\widehat A}^{2}\varphi)$ is just the norm of the vector $-i{\widehat A}\varphi$ squared. Note that the expected value $(\varphi, {\widehat A}_{\bot}\varphi)$ of the operator ${\widehat A}_{\bot} \equiv {\widehat A}-{\overline A}I$ in the state $\varphi$ is zero. Therefore, the vector $-i{\widehat A}_{\bot}\varphi=-i{\widehat A}\varphi-(-i{\overline A}\varphi)$, which is the component of $-i{\widehat A}\varphi$ orthogonal to $-i\varphi$,  is  orthogonal to the entire fibre $\{\varphi\}$.
Accordingly, the variance 
\begin{equation}
\label{varr}
\Delta A^{2}=(\varphi, ({\widehat A}-{\overline A}I)^{2}\varphi)=(\varphi, {\widehat A}_{\bot}^{2}\varphi)=(-i{\widehat A}_{\bot}\varphi, -i{\widehat A}_{\bot}\varphi) 
\end{equation}
is the norm squared of the component $-i{\widehat A}_{\bot}\varphi$. Recall that the image of this vector under $d\pi$ can be identified with the vector itself. 
It follows that the norm of $-i{\widehat A}_{\bot}\varphi$ in the Fubini-Study metric coincides with its norm in the Riemannian metric on $S^{L_{2}}$ and in the original $L_{2}$-metric. 

Furthermore, the uncertainty relation
\begin{equation}
\label{uncert}
\Delta A \Delta B \ge \frac{1}{2}\left |\left(\varphi, [{\widehat A},{\widehat B}]\varphi\right)\right|
\end{equation}
follows now from the comparison of areas of rectangle $A_{|XY|}$ and parallelogram  $A_{XY}$ formed by vectors $X=-i{\widehat A}_{\bot}\varphi$ and $Y=-i{\widehat B}_{\bot}\varphi$:
\begin{equation}
\label{obv}
A_{|XY|} \ge A_{XY}.
\end{equation}
There is also an uncertainty identity, Ref.\cite{KryukovUncert}:
\begin{equation}
\label{Pyth}
\Delta A^{2} \Delta B^{2}=A^{2}_{XY}+ G^{2}_{\varphi} (X, Y).
\end{equation}

Suppose that observables ${\widehat A}$, ${\widehat B}$ satisfy the canonical commutation relation $[{\widehat A},{\widehat B}]=i$. A unitary transformation ${\widehat U}$ preserves the commutation relation, which together with (\ref{varr}) yields
\begin{equation}
\left\|-i({\widehat U}^{-1}{\widehat A}{\widehat U})_{\bot}\varphi\right\| \left\|-i({\widehat U}^{-1}{\widehat B}{\widehat U})_{\bot}\varphi\right\|\ge \frac{1}{2}.
\end{equation}
Note that the uncertainty of the operators changes under a general unitary transformation ${\widehat U}$. In other words, the normal component of the associated vector field changes. At the same time the estimate for the product, i.e., the uncertainty relation itself, remains the same. The geometric nature of the uncertainty relation makes this invariance particularly transparent.

\section{Geometry of the Scr{\"o}dinger evolution}

Any vector in the Hilbert space, attached to a point $\varphi$ on the sphere, can be decomposed onto the radial component (parallel to the radius vector from the origin to the point $\varphi$, i.e., parallel to $\varphi$ itself), and tangential component. Furthermore, the tangential component can be decomposed onto the one parallel to the fibre (i.e., tangent to the phase circle through $\varphi$) and  the one orthogonal to the fibre. As already discussed, under the projection onto the projective space $CP^{L_{2}}$ the parallel tangential component gives zero, while the orthogonal tangential component can be identified with the vector tangent to $CP^{L_{2}}$. 

Let's see what can be extracted from these simple geometric facts. Consider the Schr{\"o}dinger equation
\begin{equation}
\label{evoll}
\frac{d\varphi}{dt}=-i{\widehat h}\varphi
\end{equation}
for the state $\varphi$ with the initial condition $\left.\varphi \right|_{t=0}=\varphi_{0}$.
Notice that 
\begin{equation}
{\mathrm Re} (\varphi, -i{\widehat h} \varphi)=0,
\end{equation}
which means that the radial component of the velocity vector $\frac{d\varphi}{dt}$ is zero. This fact is clear already from its tangential nature. Let's decompose the tangent vector $-i{\widehat h}\varphi$ onto the components parallel $||$  and orthogonal $\perp$ to the fibre. The parallel component of $\frac{d\varphi}{dt}$ is numerically
\begin{equation}
{\mathrm Re} (-i\varphi, -i{\widehat h} \varphi)={\overline E},
\end{equation}
i.e., the expected value of the energy. So the decomposition of the velocity vector $\frac{d\varphi}{dt}$ into the parallel and orthogonal components is then given by 
\begin{equation}
\label{evolll}
\frac{d\varphi}{dt}=-i{\overline E}\varphi+\left(-i{\widehat h}\varphi-i{\overline E}\varphi\right)=
-i{\overline E}\varphi-i {\widehat h}_{\perp}\varphi.
\end{equation}
We see that the orthogonal component of the velocity  $\frac{d\varphi}{dt}$ is equal to $-i {\widehat h}_{\perp}\varphi$, so we immediately conclude that: 
{\it The velocity of evolution of state in the projective space is equal to the uncertainty of energy.}
Note that (\ref{evolll}) also demonstrates that the physical state is driven by the operator  ${\widehat h}_{\perp}$, which defines the uncertainty in energy rather than the energy itself.

Now let's decompose the acceleration vector $\frac{d^{2}\varphi}{dt^{2}}=\frac{d}{dt}\left( -i{\widehat h}\varphi\right)=-{\widehat h}^{2}\varphi$. Notice first of all that
\begin{equation}
{\mathrm Re}(-i\varphi, {\widehat h}^{2}\varphi)=0,
\end{equation}
so that the parallel tangential component of acceleration of Shr{\"o}dinger evolution vanishes. This simply means that the phase component of the velocity (i.e., the expected value of energy, see above) does not change. In particular, the tangential component is purely orthogonal. The radial component is given by $-(\varphi, {\widehat h}^{2}\varphi)\varphi=-(-i{\widehat h}\varphi, -i{\widehat h}\varphi)\varphi$. Since $-i{\widehat h}\varphi$ is the velocity of evolution, we recognize in this term the centropidical acceleration ($-\frac{{\bf v}^{2}{\bf r}}{r^{2}}$ with $r=1$).

The tangential component is therefore equal to 
\begin{equation}
-{\widehat h}^{2}\varphi + (\varphi,{\widehat h}^{2}\varphi)\varphi=-{\widehat h}^{2}_{\perp}\varphi.
\end{equation}
Therefore, the following result is obtained: {\it Acceleration of the Schro{\"o}dinger evolution of state in the projective space is equal to the uncertainty of the square of energy.} Note also that the latter uncertainty can be interpreted as the fourth order central moment of the probability distribution associated with energy.

\section{Geometry of transition from quantum to classical}

Classical and quantum mechanics of a particle are now formulated within the same Hilbert space framework. Could it be that classical motion is a projection of some sort of the Schr{\"o}dinger evolution on the submanifold $M_{3}$?

One standard way to describe the relation of the Schr{\"o}dinger evolution with the classical Newtonian motion is via the Ehrenfest theorem (the expected value of the Heisenberg equation of motion):
\begin{equation}
\label{Her}
\frac{d}{dt}(\varphi, {\widehat A}\varphi)=-i(\varphi, [{\widehat A}, {\widehat h}]\varphi).
\end{equation}
Here ${\widehat A}$ does not depend on $t$. For example, for the momentum operator of a free particle we obtain
\begin{equation}
\label{para}
\frac{d{\overline{\bf p}}}{dt}=0.
\end{equation}
Recall that ${\overline{\bf p}}$ is the phase projection of the vector field $p_{\varphi}$. The equation (\ref{para}) simply says that this projection is time-independent. Note that the orthogonal projection, i.e. the uncertainty $\Delta{\bf p}$ is also preserved in this case and this is not captured in (\ref{Her}).

Compare (\ref{Her})  to another equation that follows from the Schr{\"o}dinger dynamics:
\begin{equation}
\label{projj}
 2\left(\frac{d  \varphi}{dt}, -i {\widehat A} \varphi \right)=
 \left( \varphi, \{{\widehat A}, {\widehat h} \}\varphi \right)-\left(\varphi,[{\widehat A}, {\widehat h}]\varphi\right).
\end{equation}
The Ehrenfest theorem (\ref{Her}) for a time-independent observable amounts to using  the imaginary part of (\ref{projj}). The left hand side of (\ref{projj}) is the projection of the velocity of state onto the vector field associated with the observable ${\widehat A}$. It will be now argued that the real part of the projection may be even more relevant to the problem of transition from quantum to classical.

Recall that the classical Euclidean space $\R^{3}$ is realized by submanifold $M_{3}$ in the Hilbert space  ${\bf H}\approx L_{2}(\R^{3})$ defined by (\ref{hh}). The isomorphism ${\widehat \rho}: {\bf H} \longrightarrow L_{2}(\R^{3})$ given by
\begin{equation}
\rho({\bf x},{\bf y})=\left(\frac{1}{\pi \sigma^{2}}\right)^{3/4}e^{-\frac{({\bf x}-{\bf y})^{2}}{2\sigma^{2}}}
\end{equation}
identifies $M_{3}$ with the submanifold $M^{\sigma}_{3}$ of Gaussian functions of width $\sigma$ and norm $1$ in $L_{2}(\R^{3})$. In the following it will be assumed that the length in $\R^{3}$ is measured in the units of  $\sqrt{2}\sigma$. Then, as before, the manifold $M^{\sigma}_{3}$ is isometric to the Euclidean space $\R^{3}$.


Because all normalized Gaussian functions of a given width $\sigma$ are obtained from a single one by translations in ${\bf x}$, the field ${\bf p}_{\varphi}=-i{\widehat {\bf p}}\varphi$ for $\varphi \in M^{\sigma}_{3}$ is tangent to $M^{\sigma}_{3}$. 
Consider a wave packet which is a solution of the uniform acceleration problem in potential $V$ in one dimension 
\begin{equation}
\label{theta}
\varphi(x,t)=\frac{1}{(\pi)^{\frac{1}{4}}(\sigma^{2}+\frac{t^{2}}{m^{2}\sigma^{2}})^{\frac{1}{4}}} e^{-\frac{(x-x_{0}-v_{0}t-\frac{1}{2}wt^{2})^{2}}{2(\sigma^{2}+\frac{t^{2}}{m^{2}\sigma^{2}})}}e^{i\theta},
\end{equation}
where $\theta$ is a phase factor, $v_{0}$ is the initial group velocity of the packet and $w=-\nabla V/m$. The map $\pi: \varphi \longrightarrow r$, where $r$ is the modulus of $\varphi$, projects  the initial state $\varphi_{0}(x)=\varphi(x,0)$ onto a point on $M^{\sigma}$, which is a one-dimensional version of $M^{\sigma}_{3}$. (If $\varphi$ is represented by the pair $(r,\theta)$, then $\pi(r,\theta)=(r,0)$. Physically, the phase factor plays no role in the probability of collapse to a position eigenstate. The spaces $M^{\sigma}_{3}$, $M^{\sigma}$ must be therefore identified with the set of equivalence classes of states defined up to the phase $\theta$. See below.)  The derivative map $d\pi$ projects the vector $\frac{d \varphi}{dt}$ in the tangent space $T_{\varphi}S^{L_{2}}$ onto a vector in the tangent space $T_{r}S^{L_{2}}$. The component of this vector that is tangent to $M^{\sigma}$ at $t=0$ can be obtained using the Riemannian metric on the sphere and is equal to 
\begin{equation}
\label{53}
\mathrm{Re}\left(\frac{d  \varphi}{dt}e^{-i\theta}, - \frac{d r}{dx} \right)=\left(\frac{d r}{dt}, - \frac{d r}{dx} \right),
\end{equation}
where parentheses denote the usual $L_{2}$ inner product. Note that the vector $-i{\widehat p}r=-\frac{d r}{dx}$ is unit in the chosen units. 
A direct calculation shows that at $t=0$
\begin{equation}
\label{v0}
\left(\frac{dr}{dt},-\frac{dr}{dx}\right)=-v_{0}.
\end{equation}
The left hand side of (\ref{v0}) is simply the $M^{\sigma}$-component of the velocity of the shadow $\pi(\varphi)=r$ of $\varphi$ at $t=0$.
Note that the component of $d\varphi/dt=-i{\widehat h}\varphi$ that gives $v_{0}$ via (\ref{v0}) is orthogonal to the fibre through $\varphi$, i.e., it comes from the term $-i{\widehat h}_{\perp}\varphi$. So, {\it the velocity $v_{0}$ of the shadow is due to the uncertainty $\Delta h$ in the energy. }
Note also that the component in (\ref{53}) is a part of the term
\begin{equation}
\mathrm{Re}\left(\frac{d  \varphi}{dt}, -i {\widehat p} \varphi \right)=\left(\frac{d r}{dt}, -\frac{d r}{dx} \right)-\left(r\frac{d \theta}{dt}, r\frac{d \theta}{d x} \right),
\end{equation}
which is given itself by the anticommutator term in (\ref{projj}).

Similary, the acceleration of the shadow at $t=0$ is given by
\begin{equation}
\label{a}
\left(\frac{d^{2}r}{dt^{2}},-\frac{dr}{dx}\right)=-w.
\end{equation}
These results were extended in Ref.\cite{KryukovForster}.

The equations (\ref{v0}), (\ref{a}) hint that 
the classical motion of a macroscopic particle may indeed be a projection of the Schr{\"o}dinger evolution onto $M^{\sigma}_{3}$. To prove that this is the case one must first explain what happens at $t \neq 0$. The conjecture is that the state of the particle experiences a periodic ''collapse'' so that the "width" of the state following the collapse returns to the value $\sigma$.  Suppose the state $\varphi$ undergoes the Schr{\"o}dinger evolution (\ref{theta}) between $t=0$ and the collapse episode at $t=t_{1}$. To ensure the validity of (\ref{v0}), (\ref{a}) at the moment $t=t_{1}$, suppose that the state $\psi(t_{1})$ right after the collapse is the state $\varphi(t_{1})$ given by (\ref{theta}), but with the width
\begin{equation}
\sigma^{2}+\frac{t^{2}_{1}}{m^{2}\sigma^{2}}
\end{equation}
in all terms (including phase) changed to the value $\sigma^{2}$.  More precisely, the state  $\psi(t_{1})$ is simply the result of the Schr{\"o}dinger evolution (\ref{theta}) of the initial state $\psi_{0}$, which is the original initial state $\varphi_{0}$ but with $\sigma$ replaced with the smaller value $\tilde{\sigma}$ that satisfies the equation
\begin{equation}
\tilde{\sigma}^{2}+\frac{t^{2}_{1}}{m^{2}\tilde{\sigma}^{2}}=\sigma^{2}.
\end{equation} 
The equations (\ref{v0}) (with $v(0)=v_{0}$ replaced by $v(t_{1})=v_{0}+wt_{1}$) and  (\ref{a}) are now satisfied for the state $\psi$ at $t=t_{1}$ simply because they are satisfied for the state $\varphi$ at $t=0$.

To put this in a clear geometric context, consider the fibre bundle $p: E  \longrightarrow M^{\sigma}$
over the manifold $M^{\sigma}$, whose fibre over $\tilde{\delta}_{a}\in M^{\sigma}$ consists of all normalized Gaussian functions with the mean $a$, multiplied by an arbitrary phase factor $e^{i\theta(x,t)}$. Let $G$ be the group of transformations on $E$ generated by scaling in $x$ and multiplication by phase factors $e^{i\theta(x,t)}$. 
Define the (Ehresmann) connection on $E$ by calling a tangent vector at a point $\varphi$ horizontal if it is orthogonal in the $L_{2}$ metric to the fibre through $\varphi$. In particular, the tangent space $T_{\tilde{\delta}_{a}}M^{\sigma}$ to $M^{\sigma}$ at $\tilde{\delta}_{a}$ is horizontal. The bundle projection $p$ is a transformation in $G$.  The connection is invariant under $G$, so that the vector  $-dr/dxe^{i\theta}$ is horizontal along the fibre. Moreover, the projection 
\begin{equation}
\mathrm{Re}\left(\frac{d \varphi}{dt}, -\frac{dr}{dx}e^{i\theta}\right)=\mathrm{Re}\left(\frac{d \varphi}{dt}e^{-i\theta}, -\frac{dr}{dx}\right)=\left(\frac{d r}{dt}, -\frac{dr}{dx}\right)
\end{equation}
 is constant along the fibre, so that the equation (\ref{v0}) with $-v(t)$ on the right remains true at all times.


When applied to microscopic particles, this result explains, in particular, why these particles move as if they follow a classical trajectory. Namely, the shadow $p(\varphi)$ of the state on the base $M^{\sigma}$ moves classically. Note that this result is not about the change in the expected value of the position operator. It is about the change in the state itself and how this change is "observed" on $M^{\sigma}_{3}$. 

What distinguishes macroscopic particles in this scenario is the mentioned process of collapse along the fibres of the bundle $E$. In this case the shadow of the state continues to satisfy the Newtonian equations of motion. Additionally, collapse ensures that the state remains "localized". Notice that the described collapse does not make the phase terms vanish. The classical space appears here as the set of equivalence classes of states of different width and phase.
The question of what is a possible physical origin of collapse and how the above picture is consistent with the Born rule will be discussed in sections X and XI.

The Schr{\"o}dinger equation in one dimensional case is equivalent to the following two equations:
\begin{equation}
\frac{dr}{dt}=-\frac{1}{m}\frac{dr}{dx}\frac{d \theta}{dx}-\frac{r}{2m}\frac{d^{2}\theta}{dx^{2}},
\end{equation}
\begin{equation}
\frac{d\theta}{dt}=-\frac{1}{2m}\left(\frac{d\theta}{dx}\right)^{2}-V+ \frac{1}{2mr}\frac{d^{2}r}{dx^{2}}.
\end{equation}
The first equation yields the continuity equation (the quantum story) and the second yields Hamilton-Jacobi equation (the classical story).
The obtained result suggests that the first equation may be even more relevant to transition from quantum to classical. Moreover, the equation for the state itself rather than for the associated probability density seems to be in the heart of the transition. 

\section{Quantum probability and the classical normal distribution}

If a classical experiment is set with the goal of measuring the position of a macroscopic particle, the result is generically a normal probability distribution of the position variable. Recall now that the classical space $\R^{3}$ can be identified with the submanifold $M^{\sigma}_{3}$ in the Hilbert space $L_{2}$ of states (equivalently, with the submanifold $M_{3}$ in the space ${\bf H}$). The experiment of measuring position of a macroscopic particle is then taking place within the sphere of states $S^{L_{2}}$. Because of this, the experiment has a huge impact on the possible outcomes of an experiment that measures the position of a microscopic particle.  

In particular, suppose that results of a position measurement of a macroscopic particle are normally distributed. Because $\R^{3}$ is identified with $M^{\sigma}_{3}$, the normal distribution of position dictates the corresponding probability of transition between different states of the particle in $M^{\sigma}_{3}$. 
Suppose now that the probability for the particle in an arbitrary state $\varphi$  to be found in a state $\psi$ depends only on the distance $\rho(\pi(\varphi), \pi(\psi))$ between the states, in the Fubini-Study metric on the projective space $CP^{L_{2}}$. Then the Born rule for probability of transition between the states 
follows.

The proof of this claim follows immediately from the fact that the Born rule applied to the states in $M^{\sigma}_{3}$ yields the normal probability distribution. In fact, for the state $\tilde{\delta}_{\bf a}$ in $M^{\sigma}_{3}$ given by
\begin{equation}
\tilde{\delta}({\bf x}-{\bf a})=\left(\frac{1}{\pi \sigma^{2}}\right)^{3/4}e^{-\frac{({\bf x}-{\bf a})^{2}}{2\sigma^{2}}},
\end{equation}
the probability density of finding the particle at ${\bf b}$ is equal to
\begin{equation}
\left(\frac{1}{\pi \sigma^{2}}\right)^{3/2}e^{-\frac{({\bf a}-{\bf b})^{2}}{\sigma^{2}}},
\end{equation}
which is the normal distribution function. So, on the states $\tilde{\delta}_{\bf a}$ that form the classical space $M^{\sigma}_{3}$  the quantum-mechanical Born rule makes the same prediction as the normal probability distribution. 
On the other hand, the corresponding probability of finding the particle in the state $\tilde{\delta}_{\bf b}$ is equal by the same Born rule to
\begin{equation}
\label{PP}
P(\tilde{\delta}_{\bf a}, \tilde{\delta_{\bf b}})=\cos^{2}\rho(\tilde{\delta}_{\bf a}, \tilde{\delta_{\bf b}}),
\end{equation}
where $\rho(\tilde{\delta}_{\bf a}, \tilde{\delta_{\bf b}})$ is the distance between the states $\tilde{\delta}_{\bf a}, \tilde{\delta_{\bf b}}$ in the projective space $CP^{L_{2}}$. Note that because $M^{\sigma}_{3}$ can be identified with a submanifold in the base $CP^{L_{2}}$ of the fibre bundle $\pi: L_{2} \longrightarrow CP^{L_{2}}$, we set $\pi(\tilde{\delta}_{\bf a})=\tilde{\delta}_{\bf a}$ in (\ref{PP}). Note also that the space of states here is the Hilbert space $L_{2}(\R^{3})$. The projective space $CP^{L_{2}}$ is only used in discussions related to the Fubini-Study distance between states.

Observe that an arbitrary state is a superposition of the states $\tilde{\delta}_{\bf a}$. Furthermore, the Fubini-Study distance between the states $\tilde{\delta}_{\bf a}$, $\tilde{\delta}_{\bf b}$ takes on all values from $0$ to $\pi/2$, which is the largest possible distance between points in $CP^{L_{2}}$. In addition, the probability $P(\varphi, \psi)$ to find state $\varphi$ in the state $\psi$ was assumed to depend only on the Fubini-Study distance $\rho(\pi(\varphi), \pi(\psi))$ between the states. From this and (\ref{PP}) it follows that
\begin{equation}
P(\varphi, \psi)=\cos^{2}\rho(\pi(\varphi), \pi(\psi)),
\end{equation} 
which is the Born rule for arbitrary states. In other words, the normal probability distribution for states in $M^{\sigma}_{3}$ extends to the Born rule for superpositions of these states.

This beautiful result is based on a highly non-trivial way in which the classical space is embedded into the Hilbert space of states. Namely, because of the special properties of the embedding, the "classical law" (normal distribution of observation results) becomes a part of the quantum law, which simply extends the classical law to superpositions. The extension is unique if the assumption is made that the probability of transition must only depend on the distance between states in the Fubini-Study metric. 

In more detail, assume the normal probability distribution of the position variable ${\bf a}$. Denote the distance between two points ${\bf a}, {\bf b}$ in $\R^{3}$ by $\left\|{\bf a}-{\bf b}\right\|_{\R^{3}}$. Under the embedding of the classical space into the space of states, the variable ${\bf a}$ is represented by the state $\tilde{\delta}_{\bf a}$. The set of states $\tilde{\delta}_{\bf a}$ form a submanifold $M^{\sigma}_{3}$ in the Hilbert spaces of states $L_{2}(\R^{3})$. The manifold $M^{\sigma}_{3}$ is "twisted" in $L_{2}(\R^{3})$, it belongs to the sphere $S^{L_{2}}$ and spans all dimensions of $L_{2}(\R^{3})$. Distance between the states $\tilde{\delta}_{\bf a}$, $\tilde{\delta}_{\bf b}$ in $L_{2}(\R^{3})$ or in the projective space $CP^{L_{2}}$ is not equal to $\left\|{\bf a}-{\bf b}\right\|_{\R^{3}}$. In fact, the former distance measures length of a geodesic between the states while the latter is obtained using the same metric on the space of states, but applied along a geodesic in the twisted manifold $M^{\sigma}_{3}$. In precise terms the relation between the two distances is given by
\begin{equation}
\label{main}
e^{-\frac{({\bf a}-{\bf b})^{2}}{2\sigma^{2}}}=\cos^{2}\rho(\tilde{\delta}_{\bf a}, \tilde{\delta_{\bf b}}).
\end{equation}
This equation is what accounts for the relation between the normal probability distribution and the Born rule.


\section{Collapse of the state function}

By virtue of (\ref{main}), the Born rule is an extension of the normal probability distribution from $M^{\sigma}_{3}$ onto $L_{2}(\R^{3})$.  It is therefore natural to expect that quantum measurement and collapse of the state function can be explained via the corresponding extension of the classical measurement. An attempt to build a simple extension of this kind and to draw conclusions from it will now be presented. The results of this section, although relatively straightforward, require a rigorous mathematical foundation and will be presented in detail elsewhere.

Consider an experiment of measuring position of a macroscopic particle. Suppose that interaction in the experiment depends only on the distance between particles and that the resulting distribution of position  ${\bf a}$ of the measured particle is normal.
Using the embedding $\tilde{\omega}: {\bf a} \longrightarrow \tilde{\delta}^3_{\bf a}$, we can "transplant" the model of this experiment from $\R^{3}$ onto the submanifold $M^{\sigma}_{3}$ in $CP^{L_{2}}$. Then we can extend the model to the projective space $CP^{L_{2}}$  and to the space of states $L_{2}(\R^{3})$ by replacing distance between "delta" states with the Fubini-Study distance between arbitrary states (which in particular gives correct distance between the "delta" states). By (\ref{main}), the transition probability $P(\varphi,\psi)$ in the obtained model will satisfy the Born rule.

One simple method of obtaining a normal distribution of the position of a macroscopic particle is by the process of diffusion of the particle in an appropriate medium (Brownian motion). Suppose a macroscopic particle is placed in the medium and position of the particle a certain time interval $\tau$ later is recorded.  The probability density function of the position variable is determined by the diffusion coefficient $D$. The process is isotropic: the coefficient $D$ depends only on the frequency of various observed displacements of the particle in the medium, but not the direction of these displacements. Therefore, diffusion can be extended to a process on the projective space $CP^{L_{2}}=S^{L_{2}}/S^{1}$, with the diffused particle realized now by the state $\{\varphi\} \in CP^{L_{2}}$ and with displacements in $\R^{3}$ replaced by the like-displacements of the state in the Fubini-Study metric. As discussed, the probability of finding the state $\{\varphi\}$ of the "diffused" particle in a state $\{\psi\}$ (i.e., at the point $\{\psi\} \in CP^{L_{2}}$) is then given by the Born rule:
$P(\{\varphi\}, \{\psi\})=\cos^{2}\rho(\{\varphi\}, \{\psi\})$.

Note that the obtained process, which we still call diffusion, or diffusion in the space of states, gives correct probabilities of outcomes of measurement of {\it any} observable. In fact, the eigenstates of observables are just points in the space of states. The diffusion can take the particle to an arbitrary eigenstate and the probability of the resulting transition is given by the Born rule. In this sense diffusion in the space of states serves as a model for an arbitrary quantum measurement. Suppose, for example, that we want to find the probability density of various positions of a particle. To measure position, we subject the particle (i.e., the state) to diffusion in the space of states. If in the process of diffusion the particle reaches the manifold $M_{3}$, then position ${\bf a} \in \R^{3}$ of the particle becomes meaningful and we say that collapse of the state has occurred. Note that the probability density function of finding the particle at ${\bf a}$ is the {\it conditional} probability density, given that the diffusive particle  was found on $M_{3}$. 

In this model the process of collapse is independent of a particular measurement performed on the particle. It is simply a random walk that leads to an eigenstate of the measured observable. In the case of position measurement the initial state simply happened to be "pushed" in the space of states in the direction of decreasing width $\sigma$, i.e., towards the submanifold $M^{\sigma}_{3}$. 

Of course, collapse by diffusion does not preclude other forms of collapse. In fact, any process that involves a sufficient "squeezing" of the state in the position representation can be thought of as collapse to a point in $\R^{3}$. In particular, an appropriate potential well that narrows down in time will drive the state to the submanifold $M_{3}$ thus providing a model of controlled collapse to a position eigenstate. This latter example together with the Schr{\"o}dinger equation itself prove that our usual instruments are capable of accessing points (i.e., states) outside the classical space $M_{3}$. In other words, the idea that a classical measuring device is responsible for a physical process on the space of states, and not just the classical space, seems to be realized in nature. Note also that because the sphere of normalized states can be arbitrarily small, the extension of a classical process to the space of states may be limited to an arbitrarily thin layer around the submanifold $M_{3}$. Furthermore, vector fields $dr/d\sigma$ and $dr/dx$ are orthogonal, showing that collapse has nothing to do with motion in the classical space (see Sec.IX). In particular, it does not make sense to impose a speed limit in the classical space on this process.

The introduced collapse by diffusion sheds new light into the relationship between classical and quantum measurements. If position of a macroscopic particle is measured, the diffusion process models a noise present in any measurement and the resulting normal probability distribution of measurement results. In this case the position of the particle at time $\tau$ can be observed no matter where the particle is located. On the other hand, in the experiment of measuring position of a microscopic particle, the position is not readily available. The state at time $\tau$ (i.e., the end-point of the random walk) does, of course, exist. But position in the usual sense is meaningful only when diffusion takes the state to the classical space $M_{3}$. 

The situation can be elucidated by an example from the classical physics in which position measuring devices are on the $X$-axis in the Euclidean space $\R^{3}$. If this represents the only position measurement available to us, then a macroscopic particle has a definite position only if it is on the $X$-axis. The latter situation can be achieved, in particular, by subjecting the particle to diffusion in $\R^{3}$. When the drift motion in the diffusion can be neglected, the expected value of the measured position is simply the $x$-coordinate of the particle at the moment of measurement.


It was shown that diffusion can explain the outcomes of a measurement and the Born rule. Moreover, it is also capable of explaining why the state function of a macroscopic body does not  spread. For this note first of all that a macroscopic body interacts continuously with the environment. This interaction is similar to the bombardment of a Brownian particle by molecules of the medium. However, because of the macroscopic character of the body, the diffusion process does not occur and position of the body remains unaffected by the interaction (e.g., a boat initially at rest in still water will not change its position). If now the interaction is extended to the space of states, then the conclusion is that the state of  a macroscopic particle will not be the subject of a random walk. Since the increase in width is the increase in the distance from the state to the space $M_{3}$, we conclude that the width of the state function will not change.

Collapse as a physical process was 
 discussed by several authors (see Ref.\cite{Bassi} for a review).  In these works collapse is introduced by hand, either via a term in the stochastic Schr{\"o}dinger equation or via random hits (consisting in multiplication by a narrow Gaussian function) that the state function of the particle undergoes. Even though no specific reference to the Fubini-Study metric is made in these models, it is implicitly present in the relationship between the norm of state-function and the probability of collapse.  The dynamics of state under collapse in the existing models depends on the kind of measurement performed on the particle. This seems to be a negative feature of the models. As argued in Ref.\cite{AdlerHorwitz}, collapse to energy eigenstates via the stochastic equation in the projective space of states proposed in Ref.\cite{Hughston} may be sufficient to interpret an arbitrary quantum measurement. However, the proposed method may provide potentially a more appealing and physically motivated solution.





\section{Conclusion and outlook}

It was proved that the theories of classical and quantum mechanics can be considered within a single functional framework. The standard quantum mechanics in the framework is simply an extension of the classical mechanics to the space of states. The unusual properties of quantum mechanics are rooted in the  strange way in which the classical space is embedded into the space of states. The classical mechanics is recovered from the condition that macroscopic bodies are constrained to the manifold $M_{3}$.  The latter condition is supported by experiment and is likely to be due to interaction between macroscopic bodies and the environment. Radiation, other surrounding particles bombard a macroscopic body and subject it to diffusion in the space of states. However, as discussed in Sec.XI, the macroscopic character of the body forestalls the diffusion and doesn't let the state leave the classical space $M_{3}$. As a result, position of the body is either defined precisely or normally distributed with a small standard deviation $\sigma$.

Contrary to this, free microscopic particles interact only occasionally with the environment and their state function can have any width. However, when a position measuring device is turned on, a microscopic object undergoes a bombardment similar to the one affecting a macroscopic body. Due to the microscopic nature of the object, its state does not remain static, but experiences a diffusion in the Hilbert space. Surrounding particles play billiard game with the object, and the rules of the game (i.e., the Born rule for collapse onto $M_{3}$) were shown to agree with the rules of the usual classical billiard.

The embedding of the classical space into the Hilbert space of states brings a totally new perspective into the observed phenomena and a new interpretation of the familiar experiments. 
Under the embedding the state gets "promoted" from a statistical notion to a physical concept that generalizes and replaces the notion of a material point. When the state is near the submanifold $M_{3}$ in the Hilbert space ${\bf H}$, the particle (i.e., the state!) behaves classically. 
When a particle goes "through" a screen with two slits, its state becomes, roughly speaking, a superposition of delta functions, which means that the particle is no longer a point of $M_{3}$. (Recall that the manifold $M_{3}$ is "made of" the delta functions, but not their nontrivial superpositions.) The entire process of passing "through" the slits is the process of refraction, when the path of the particle is "kicked out" of  the classical space $M_{3}$ (while staying in the Hilbert space of states), and then returns to $M_{3}$ when position of the particle is measured. 





It is rather straightforward to generalize the results to quantum systems of more than one particle. In the Hilbert space ${\bf H}^{n}=\otimes_{i=1}^{n}{\bf H}$ of states of an $n$-particle system, the states of non-interacting particles belong to the submanifold $D$ of the product states. The state of a system of $n$ classical particles is represented in $D$ by the product of delta states. The set of all products of delta states forms a submanifold $M_{3n}$ of $D$, which is diffeomorphic to the configuration space $\R^{3n}$. In the case of $n$ non-interacting microscopic particles, the Schr{\"o}dinger evolution of particles drives the state of the system out of the submanifold $M_{3n}$, although the path of state still takes values in $D$.
Entanglement amounts to the state of the system leaving the manifold $D$ and propagating in ${\bf H}^{n}$. If only two particles get entangled, the state of the system is the product of states of other particles, multiplied by the state of the entangled pair.  
This allows one to treat the state of the pair independently of other particles. 

The obtained results are encouraging and appealing. However, the future work will show if these results lead in fact to a consistent theory of classical mechanics, relativity and quantum theory.


\begin{thebibliography}{99}

\bibitem{KryukovIJMMS}
A. Kryukov, {\it Int. J. Math. \& Math. Sci.} {\bf 14}, 2241 (2005)

\bibitem{KryukovJMP1}
------ {\it J. Math. Phys.}  {\bf 49}, 102108 (2008) 

\bibitem{KryukovJMP2} 
------  {\it J. Math. Phys.} {\bf 59}, 022110 (2010)



\bibitem{KryukovUncert}
------ {\it Phys. Lett. A} {\bf 370}, 419 (2007)


\bibitem{KryukovFOP1} 
------ {\it Found. Phys.} {\bf 37}, 3 (2007)


\bibitem{KryukovFOP}
------  {\it Found. Phys.} {\bf 41}, 129 (2011)

\bibitem{Stu1}
E.C.G. Stueckelberg, {\it Helv. Phys. Acta} {\bf 14}, 372, 585 (1941); {\it Helv. Phys. Acta} {\bf 15}, 23 (1942)


\bibitem{HorPir}
L.P. Horwitz and C. Piron, {\it Helv. Phys. Acta} {\bf 46}, 316 (1973)

\bibitem{Hor}
L.P. Horwitz and Y. Rabin, {\it Lett. Nuovo Cimento} {\bf 17}, 501 (1976)

\bibitem{Hor2}
L.P. Horwitz, {\it Phys. Lett. A} {\bf 355}, 1 (2006)

\bibitem{HorRot}
L.P. Horwitz and F. Rotbart, {\it Phys. Rev. D} {\bf 24}, 2127 (1981)

\bibitem{Floquet}
H.L. Cycon, R.G. Froese, W. Kirsch, and B. Simon, {\it Schr{\"o}dinger Operators
with Application to Quantum Mechanics and Global Geometry}, Springer-Verlag, New York (1987)

\bibitem{Holand}
J.S. Howland, {\it Indiana Univ. Math. Jour.} {\bf 28}, 471 (1979)

\bibitem{KryukovForster} 
M. Forster and A. Kryukov, forthcoming


\bibitem{Bassi}
A. Bassi, G. Ghirardi, {\it Phys. Rep.} {\bf 379}, 257 (2003)

\bibitem{AdlerHorwitz}
S. Adler and L.P. Horwitz, {\it J. Math. Phys.} {\bf 41}, 2485 (2000)

\bibitem{Hughston}

L. P. Hughston, {\it Proc. Roy. Soc. Lond.} {\bf A 452}, 953 (1996)

\end{thebibliography}
\end{document}